\documentclass[a4paper,12pt]{article}
\usepackage{epsf,multicol,ifthen}
\usepackage[dvips]{graphicx}
\usepackage{amsmath}
\usepackage{cite}
\usepackage{hyperref}
\newcommand{\bec}{\begin{center}}
\newcommand{\ec}{\end{center}}
\newcommand{\bee}{\begin{equation}}
\newcommand{\ee}{\end{equation}}

\title{K-group identification of supergravity solutions}

\author{T.V. Obikhod\thanks{E-mail: obikhod@kinr.kiev.ua}, I.A. Petrenko\\
\small\emph{Institute for Nuclear Research, National Academy of Science of Ukraine} \\ 
\small\emph{47, prosp. Nauki, Kiev, 03028, Ukraine}}

\begin{document}
\maketitle

\abstract{The problem with Bekenstein-Hawking entropy of black hole can be resolved with quantum gravity theory with  Dp-branes as supergravity solutions of type IIB string theory. Dp-brane solutions of type IIB are a direct analog of the Schwarzschild charged hole, so called black p-branes. The coincidence of the black p-brane metrics and ten-dimensional metrics of N-parallel D3-branes was used from the viewpoint of the Azumaya structure on D-branes connected with deformation of the classical moduli space. Applying Rosenberg theorem we can classify Hilbert spaces of N coinciding Dp-branes as vector bundles through K-functor. } \\ 
\vspace*{3mm}\\
{\bf Key words}: Black holes $\cdot$ C*-algebra $\cdot$ Dixmier-Douady invariant $\cdot$ K-group $\cdot$ Supergravity. 

\newpage
\section{Introduction}
The discovery of the gravitational waves \cite{1.} has stimulated interest to the question of gravitational collapse. In General Theory of Relativity, gravity is connected with the curvature of space-time. As the mass changes the curvature also changes, which leads to gravitational waves. Scientists have demonstrated the existence of these waves through the observation of the merger of black holes as a possible way of observing the very early Universe \cite{1.}. 

	As known, \cite{2.}, gravity determines the causal structure of the Universe, as it determines causally related events of space-time to each other. As large amount of matter can be concentrated in the same region, light could be dragged back inwards, as was stressed by Laplace in 1798. This suggests the singularity in space-time where our present laws of physics breakdown. Observations of the microwave background imply the existence of a singularity in the past as the beginning of the Universe, Fig.1. 
	
\bec
{\includegraphics[width=0.35\textwidth]{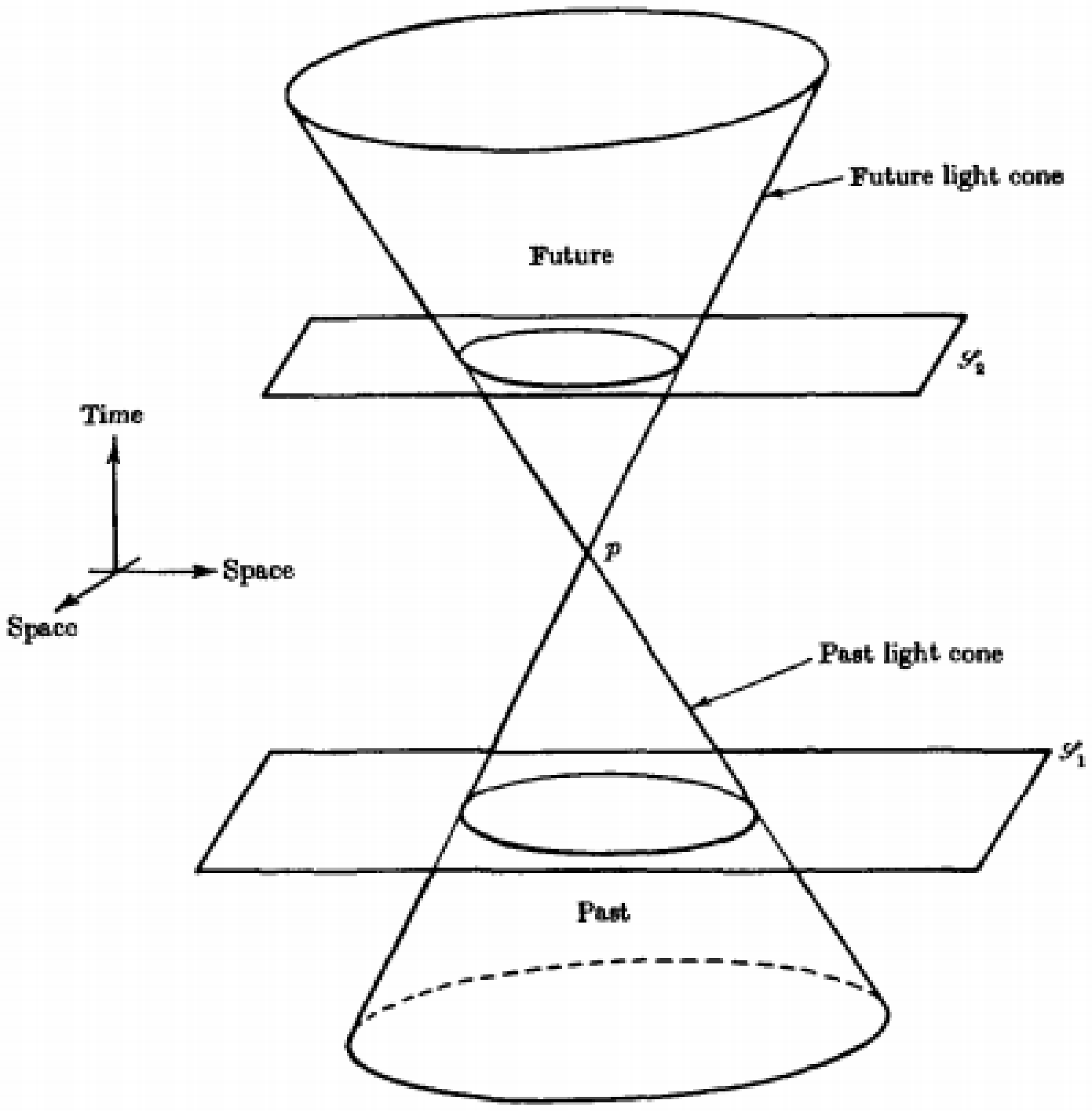}}\\
\emph{{Fig.1.}} {\emph{The light cone of the event p as the set of 
all light rays through p.}}\\
\ec

So, General Relativity predicts the development of singularities. It has been suggested in some papers \cite{3.} that after the collapse of a mass to the Schwarzschild radius, the enormous amounts of energy in the form of gravitational radiation can be emitted. The Schwarzschild radius corresponds to the radius defining the event horizon of a Schwarzschild black hole. The fact is that the black hole mass, M, decreases after Hawking radiation, proposed by Hawking for black holes in quantum mechanics \cite{4.}, and so does the area of the horizon, A$_H$, but the Bekenstein-Hawking entropy, S, which is proportional to area,
 $S=\frac{A_H}{4G_N\hbar}$, increases. Therefore, the mass of black hole, according to formula, $dM=T_HdS$, increase. So, the existence of a singularity presents problems for complete description of the physics beyond the event horizon. To overcome this puzzle was proposed quantum gravity theory with the degrees of freedom which give rise to the entropy through microstates \cite{5.}. 
The microscopic degrees of freedom provide a microscopic quantum description of extremal (M=Q for the mass of a black hole, M, in terms of its charge, Q) and near extremal (M-Q$\ll$Q) charged black holes. Further, the Dp-branes of the superstring theory can in fact be considered as supergravity solutions, which are a
direct analogue of the Schwarzschild black hole. The observables are the same for black holes as for the D-brane configurations, which can be computed in string perturbation theory. The critical black p-brane, like the Dp-branes of superstring theory, has a mass equal to the charge. Due to the mass (and charge), the p-brane bends the geometry, which can be found by solving the joint system of Maxwell-Einstein equations. Namely, the metrics looks like this: 
\begin{equation}
ds^2=\frac{1}{\sqrt{H(r)}}\Biggl( -dt^2+\sum\limits_{i=1}^p dx^idx^i \Biggr) +
\sqrt{H(r)}\sum\limits_{a=1}^{9-p} dr^adr^a
\end{equation}
The Bekenstein-Hawking entropy associated to a macroscopic black hole is explained through the condensate of closed string states with certain D-brane configurations at strong coupling, which curves space-time (metrics (1)) and may collapse to black hole space-time, Fig. 2.

\bec
{\includegraphics[width=0.7\textwidth]{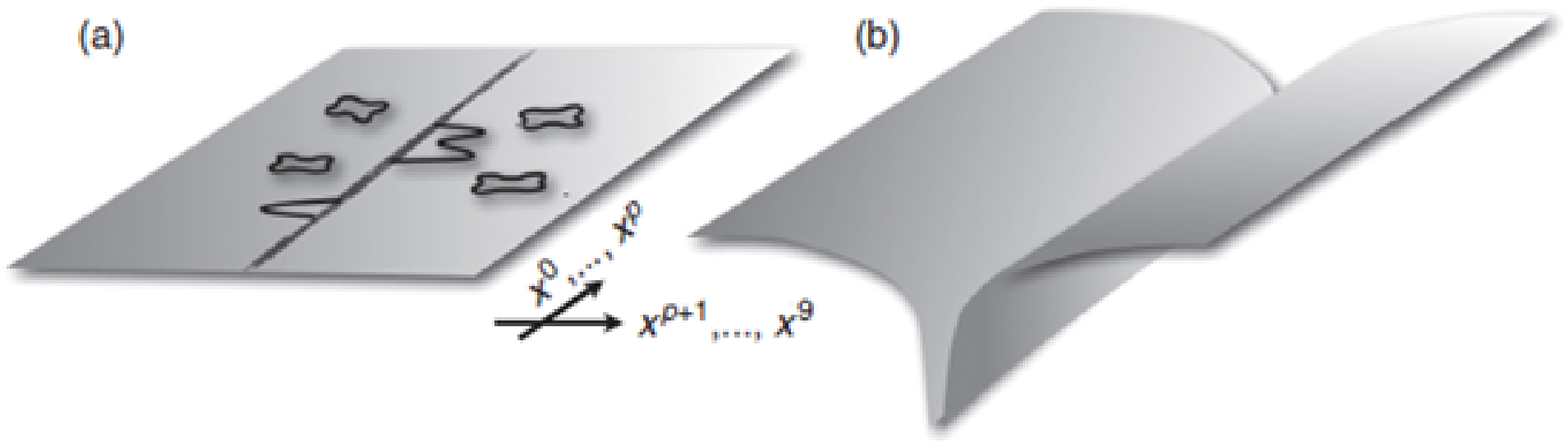}}\\
\emph{{Fig.2.}} {\emph{A Dp-brane interacts with closed strings via open strings (a), creating an effective supergravity background (b), from \cite{6.}.}}\\
\ec

\section{Computations}
The computation of entropy through the number of string states \cite{7.} is the same one as the Bekenstein-Hawking entropy of the supersymmetric black hole \cite{5.}.
It is necessary to emphasize the fact that black hole entropy can be represented by different states to be counted \cite{8.}. For the extreme black holes, one counts all BPS states in the weakly coupled string theory with the given charges. For the nearly extremal black holes, the counted states are just non-BPS excitations of the D-branes, which form a black hole at strong coupling.

	The stable non-supersymmetric Dp-branes or non-BPS states are understood as brane-antibrane system presented as Chan-Paton vector bundles, \cite{9.}. One of the most interesting cases is the identification of Dp-brane in the presence of Neveu-Schwarz B-field. Let’s consider the following two principal bundles over compact manifold X, which characterizes D-branes \cite{10.}.
\bec
\hspace*{5cm}{\includegraphics[width=0.57\textwidth]{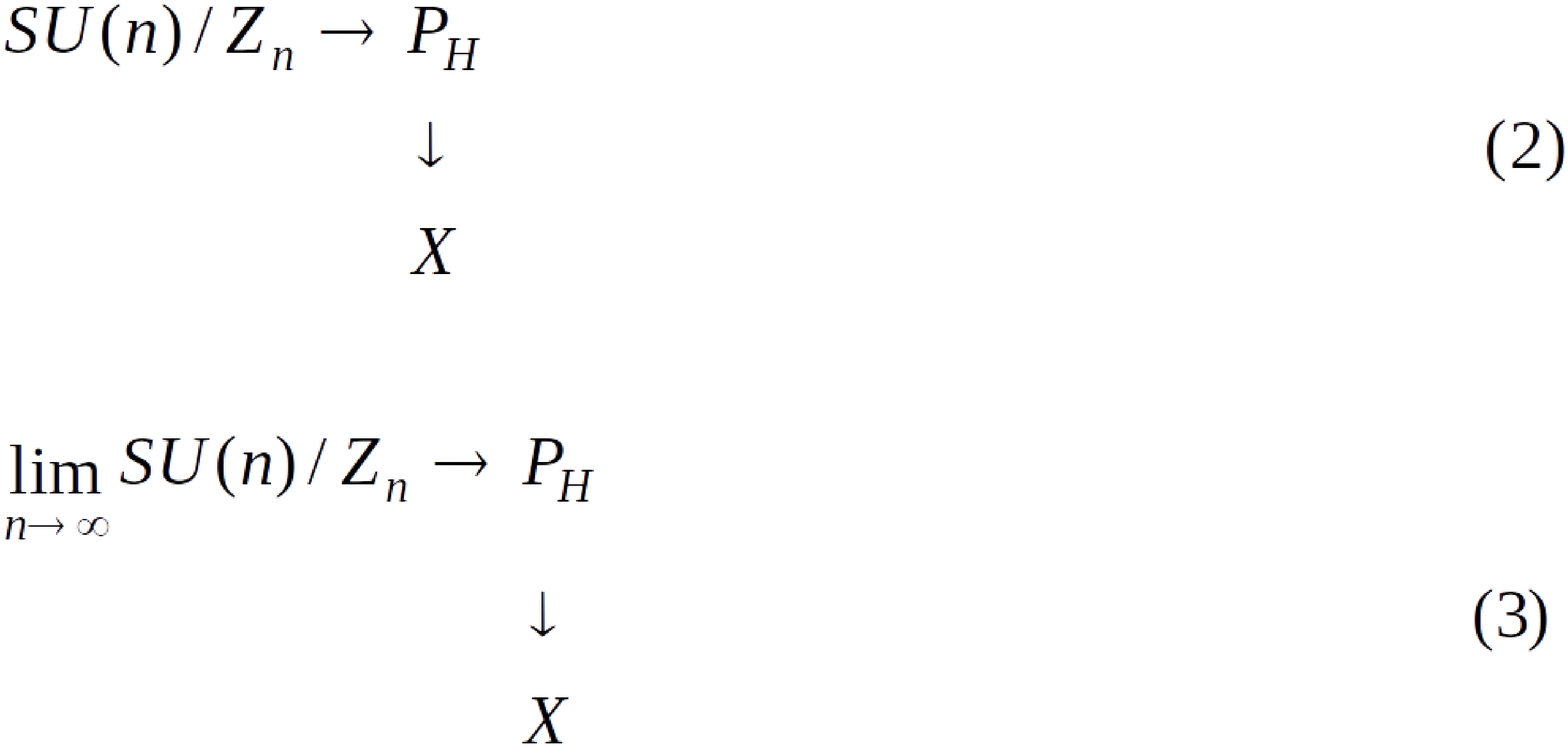}}\\
\ec
These bundles describe D-branes in the presence of Neveu-Schwarz B-field which is characterized by field strength,
\[H_{\mu\nu\rho}=\partial_{\mu}B_{\nu\rho}+\partial_{\nu}B_{\rho\mu}+\partial_{\rho}B_{\mu\nu} .\]
For the bundle (2) 
\[H_{\mu\nu\lambda}=0, \ \ \ B_{\mu\nu}\neq 0\]
 and for bundle (3) 
\[H_{\mu\nu\lambda}\neq 0, \ \ \ B_{\mu\nu}\neq 0.\]
Vector bundles associated with principal bundles are the following \cite{10.}:
\begin{equation} \tag{4}
E_H=P_H\times M_n(C), \ \ \ Aut(M_n(C))=SU(n)/Z_n \ ,
\end{equation}
\begin{equation} \tag{5}
E_H=P_H\times K, \ \ \ Aut(K)=lim_{n\rightarrow \infty}SU(n)/Z_n \ ,
\end{equation}
where $M_n(C)$ is algebra of $n\times n$ matrices and $K$ is algebra of compact operators.  The space of sections of a vector  bundle (4) is C*-algebra $C_0(X, E_H)$, which  is  Morita   equivalent  to  Azumaya algebra \cite{11.} and the space of  sections of  a  vector  bundle (5)  is  Rosenberg C*-algebra $C_0(X, E_H)$ , \cite{12.}.  Short exact sequence
\begin{equation} \tag{6}
0\rightarrow K \rightarrow B \rightarrow C_0(X, E_H) \rightarrow 0
\end{equation}
determines the extension of algebra $C_0(X, E_H)$  by means of algebra $K$, \cite{13.}. Then twisted K-group 
$K_j(X,[H])$, is determined through the set of unitarily equivalent classes of C*-algebra $C_0(X, E_H)$  extensions (6) modulo splitting extensions, \cite{14.}.
	
	As is known from \cite{15.}, for locally compact Hausdorff space, $X$ (the base of principal bundle), there exists the classifying space of the third cohomology group of $X$ - the Eilenberg-Maclane space $K(Z, 3)$, that is
\begin{equation} \tag{7}
H^3(X, Z)=[X, K(Z, 3)]=[X, PU(H)]
\end{equation}
Here,
\[K(Z, 3)= PU(H), \ \ (PU(H)=U(H)/U(1))\]
is classifying space of the projective unitary group on an infinite dimensional, separable, Hilbert space, $H$.  From \cite{15.}, $PU(H)=Aut(K)$  (group of automorphism of $K, T\rightarrow gTg^{-1}$  for $g\in U(H)$), therefore, the isomorphism classes of locally trivial bundles over $X$ with fibre $K$ and structure group $Aut(K)$ are also parametrized by $H^3(X, Z)$. The cohomology class in $H^3(X, Z)$ associated to a locally trivial bundle $E$ over $X$ with fibre $K$ and structure group $Aut(K)$ is called the Dixmier-Douady invariant of $E$, $\delta(\varepsilon)$, \cite{16.}.

	In analogy with magnetic charge of electromagnetism presented by a class in ordinary cohomology, D-brane charge is given by twisted K-theory - an additive category of C*-algebras with standard presentation by Hilbert bimodules, \cite{17.}. C∗-algebra is algebra of compact operators, $K$, on a complex, infinite-dimensional separable Hilbert space, $H$, \cite{18.}. Summarizing previous arguments, we can state that D-brane charges in the presence of $B$-field, are classified by the twisted K-theory defined by Rosenberg [12], which is the theory of infinite-dimensional, locally trivial, algebra of compact operators as introduced by Dixmier and Douady. The relevance of Dixmier-Douady theory to the classification of D-brane charges in the presence of B-fields was noticed by \cite{14.}.
	 
	For the calculation of D-brane charges in the presence of $B$-fields we will consider a configuration containing $n$ D-branes and $n$ anti-D-branes. When $[H] = 0$, the D-branes carry a principal bundle with structure group 
$U(n)$,
\bec
{\includegraphics[width=0.13\textwidth]{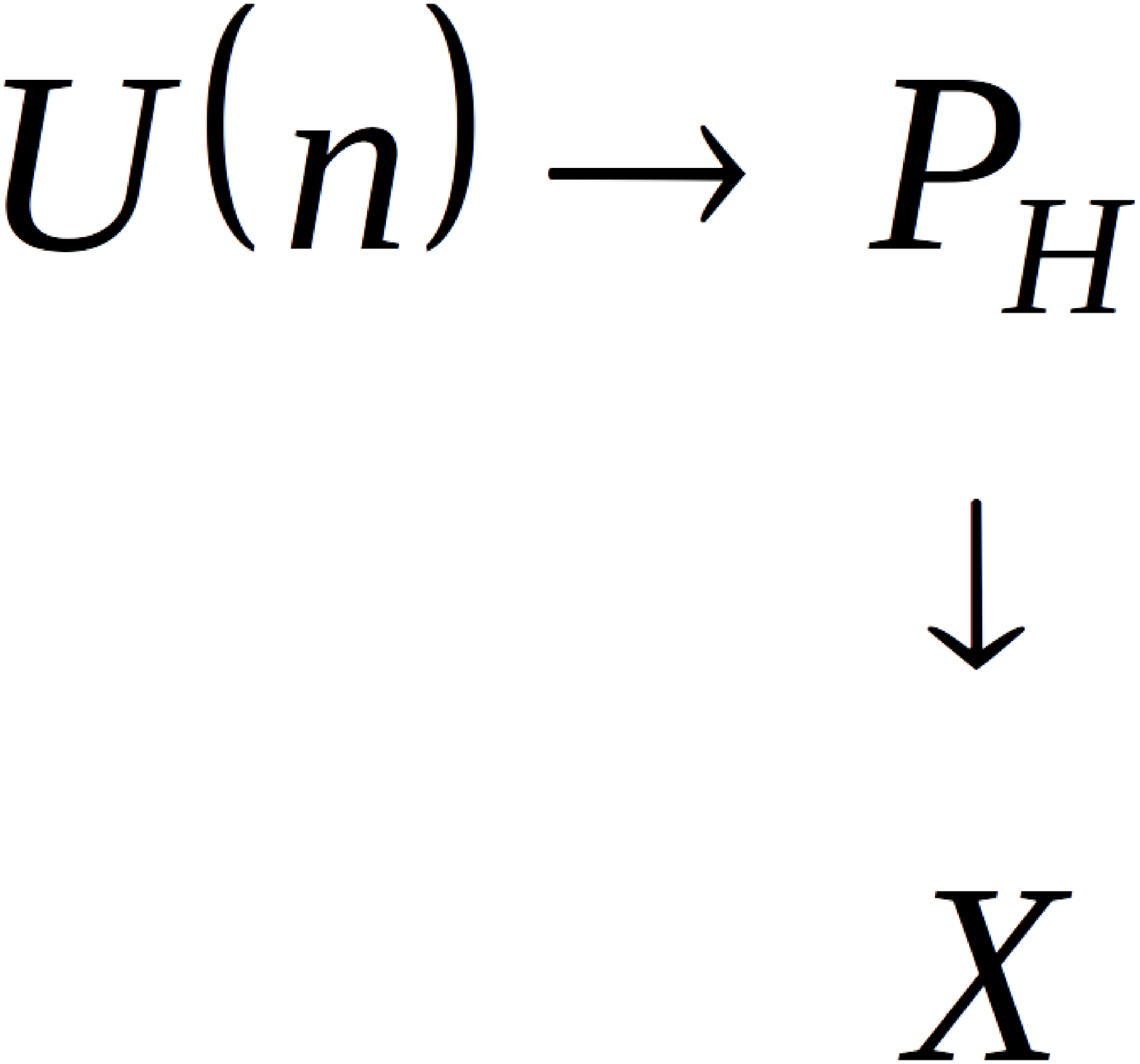}}\\
\ec
but when $[H]$ is nonzero, D-branes carry a principal $SU(n)/Z_n$ bundle which can’t be transformed to $U(n)$ bundle. The obstruction to lifting an $SU(n)/Z_n$ bundle to a $U(n)$ bundle is a certain class in $H^3(M, Z)$ closely related to the t’Hooft magnetic flux which is equal to torsion class $[H]\in H^3(M, Z)$ - a unique equivalence class of Azumaya or Rosenberg algebra and the corresponding to K-group \cite{12.}. 

	Let’s consider $\overline{X}=X/\tau$ with involution $\tau$ and principal bundle with base $\overline{X}$
\bec
{\includegraphics[width=0.11\textwidth]{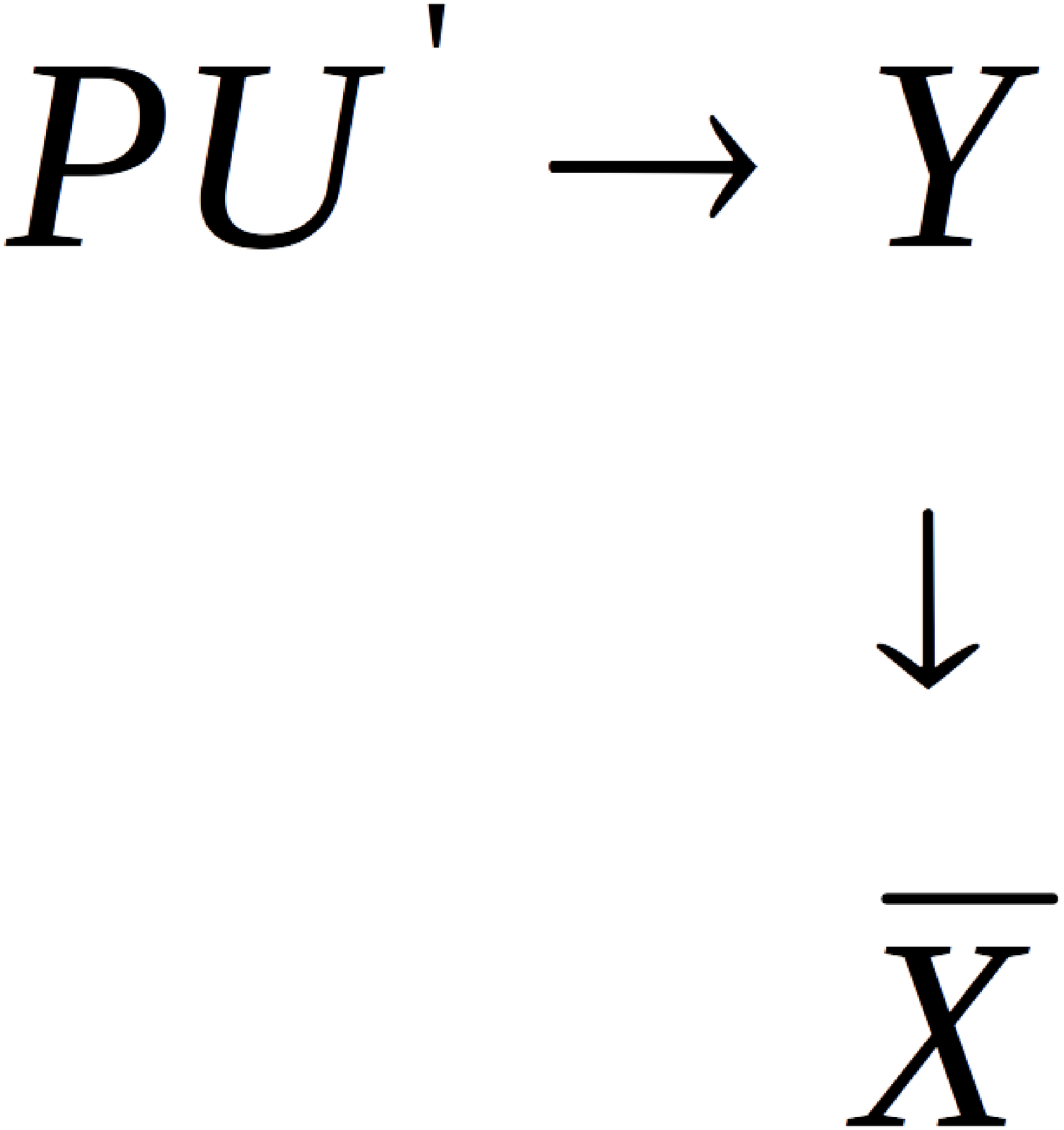}}\\
\ec

Then we have locally trivial bundles of real C*-algebras over $\overline{X}$ with fibers isomorphic to $K$. To the space of sections of vector bundles are associated principal bundles $PU^{'}$ with section - the group of unitary operators on $H$ and there are group extensions,
\[1\rightarrow PU \rightarrow PU^{'} \rightarrow Z_2\rightarrow 1\]
where $Z_2$ is the cyclic group of two elements. There exists the classifying element for $PU^{'}$-bundle $Y\rightarrow \overline{X}$  in $H^1(\overline{X}, PU^{'})=[\overline{X}, PU^{'}]$  maps to the classifying element of the two-fold covering $p: X\rightarrow \overline{X}$ defined by $\tau$ in  $H^1(\overline{X}, Z_2)\cong [\overline{X}, Z_2]$  and classifying element for principal $PU$-bundle - the Dixmier-Douady class $\delta \in [X, PU]\cong H^3(X, Z)$ defined by the pull-back diagram, \cite{12.}
\bec
{\includegraphics[width=0.14\textwidth]{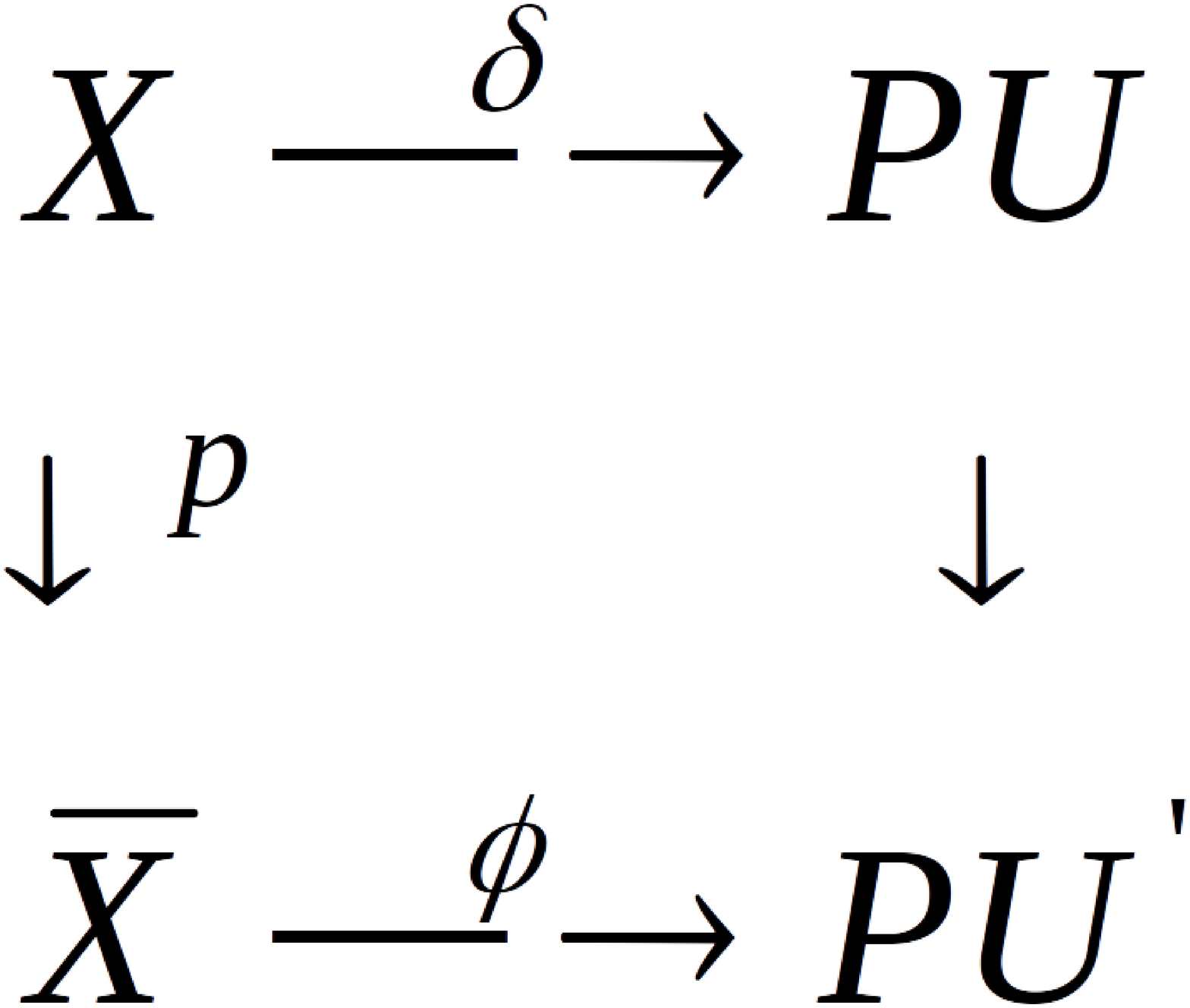}}\\
\ec
The statement of the theorem is connected with the map from one principal $PU^{'}$- bundle to another one, $PU$-bundle trough the classifying element $[\overline{X}, PU^{'}]$ which maps to classifying element of two-fold covering 
$X \rightarrow \overline{X}, [\overline{X}, Z_2]$. Therefore, we can consider bundles over  compact space, $X$ and over space $X/Z_2$ and morphisms between these bundles.

\section{Conclusions}
Since D-branes are characterized by a Dixmier-Douady invariant, to which correspond the strength of the Neveu-Schwarz $B$-field interacting with the D-branes, the transition between the classifying elements leads to new D-brane states described by a twisted K-theory. This relation allows us to associate the bundles of C*- algebras, associated with principal bundles, the observed states of Hilbert spaces, which corresponds to a phase transition from one soliton state of one D-brane to another.

	As each of them is characterized by its Dixmier-Douady invariant, which signals about the interaction of D-branes with the Neve-Schwartz $B$-field and the sections of associated vector bundles are the Hilbert spaces describing microstates of black holes, therefore, the transitions between the principal bundles are phase transitions between black holes. For an oriented tree-dimensional manifold $X$, according to the Rosenberg theorem, there is a classifying element – Dixmier-Douadi invariant and isomorphism, $\beta : H^2(X, Z_2)\rightarrow H^3(X, Z)$, which is connected with the Bockstein homomorphism 
\[0 \rightarrow Z \rightarrow Z \rightarrow Z_2 \rightarrow 0 \]
This sequence signals about new brane-antibrane bounded state $(Dp-\overline{Dp})$  for any $p$, which is classified by K-group
\[K(Dp-\overline{Dp})=Z_2\ .\]

\label{page-last} 
\label{last-page}

\newpage

\begin{thebibliography}{99}
\bibitem{1.}Abbott, Benjamin P.; et al. (LIGO Scientific Collaboration and Virgo Collaboration) (2016). Observation of Gravitational Waves from a    
Binary Black Hole Merger. Phys. Rev. Lett. 116 (6): 061102. arXiv:1602.03837.
\bibitem{2.} S. W. Hawking, and G. F. R. Ellis, The large scale structure of space-time (Cambridge University Press 1994, 391 p.).
\bibitem{3.} Ya. B. Zel'dovich and I. D. Novikov, Emission of gravitational waves by bodies travelling in the field of a collapsing star, Dokl. Akad. 
Nauk SSSR 155:5, 1964, 1033-1036. 
\bibitem{4.} S. W. Hawking, Black holes and thermodynamics, Phys. Rev. D13, 1976, 191-197.
\bibitem{5.} Juan M. Maldacena, Black Holes and D-branes, Nucl.Phys.Proc.Suppl. 61A, 1998, 111-123.
\bibitem{6.} L. Ibáñez and A. Uranga, An Introduction to String Phenomenology (Cambridge University Press 2012).
\bibitem{7.} A. Strominger and C. Vafa, Microscopic Origin of the Bekenstein-Hawking Entropy, (arXiv:hep-th/9601029).
\bibitem{8.} G. Horowitz, The Origin of Black Hole Entropy in String Theory, arXiv:gr-qc/9604051.
\bibitem{9.} E. Witten, D-branes and K-theory, JHEP 12, 1998, 019, hep-th/9810188.
\bibitem{10.} P. Bouwknegt and V. Mathai, D-branes, B-fields and twisted K-theory, hep-th/0002023
\bibitem{11.}A. Kapustin, D-branes in a topologically nontrivial B-field, hep-th/9909089
\bibitem{12.} J. Rosenberg, Continuous trace algebras from the bundle theoretic point of view, Journ. Austr. Math. Soc. (Series A) 47, 1989, 368-381.
\bibitem{13.} V. Mathai and I.M. Singer, Twistedk-homology theory, twisted Ext-theory, hep-th/0012046.
\bibitem{14.} Yu.M. Malyuta, Nonlinear problems and homological algebra, Nonlinear boundary problems, Issue 13, 2003, p. 114-117.
\bibitem{15.} P. Bouwknegt, V. Mathai, D-branes, B-fields and twisted K-theory, JHEP 0003:007,2000.
\bibitem{16.} J. Dixmier and A. Douady, Champs continues d’espaces hilbertiens at de C∗-algebres, Bull. Soc. Math. France 91, 1963, 227–284.
\bibitem{17.} Thomas Schick, L2-index, KK-theory, and connections, New York J. Math. 11, 2005, 387—443, arXiv:math/0306171.
\bibitem{18.} B. Blackadar, K-Theory for Operator Algebras (2nd ed. Cambridge University Press, Cambridge (1999)).


\end{thebibliography}
\end{document}